# A Stochastic Weather Model: A Case of Bono Region of Ghana.

Bernard Gyamfi[1*]

[1]*Postgraduate Student, Department of Statistics and Actuarial Science, College of Science, Kwame Nkrumah University of Science and Technology, Kumasi, Ghana.*

The paper sought to fit an Ornstein–Uhlenbeck model with seasonal mean and volatility, where the residuals are generated by a Brownian motion for Ghanaian daily average temperature. This paper employed the modified Ornstein–Uhlenbeck model proposed by Bhowan which has a seasonal mean and stochastic volatility process. The findings revealed that, the Bono region experiences warm temperatures and maximum precipitation up to 32.67°C and 126.51mm respectively. It was observed that the Daily Average Temperature (DAT) of the region reverts to a temperature of approximately 26°C at a rate of 18.72% with maximum and minimum temperatures of 32.67°C and 19.75°C respectively. Although the region is in the middle belt of Ghana, it still experiences warm(hot) temperatures daily and experiences dry seasons relatively more than wet seasons in the number of years considered for our analysis. Our model explained approximately 50% of the variations in the daily average temperature of the region which can be regarded as relatively a good model. The findings of this paper are relevant in the pricing of weather derivatives with temperature as an underlying variable in the Ghanaian financial and agricultural sector. Furthermore, it would assist in the development and design of tailored agriculture/crop insurance models which would incorporate temperature dynamics rather than extreme weather conditions/events such as floods, drought and wildfires.

## I. INTRODUCTION

The dynamics of the daily average temperature have acquired appeal in the research fields over the past few decades due to the concepts of global warming and the greenhouse effect combined with the world's growing population (Gao *et al.*, 2022). It is well known that the weather has a significant impact on several economic sectors, including the production of energy and the agriculture sector, weather derivatives have grown into an intriguing and new phenomenon in the field of derivatives used to hedge risk and secure against variations in the weather caused due to the aforementioned factors. With regards to agricultural industry, farmers all over the world are faced with numerous decisions which has significant impact on their performance right from preparing the farmland for plantation up until delivering their farm produces to the market. Most of these decisions are in the hands of the farmers, however, there are other existential factors that affect the yield of the farmers which is entirely out of the control of the farmers. Despite being outside of farmers' control, environmental factors including weather and climate significantly impact the performance of the farmers. It is widely acknowledged that the global climate keeps alternating, and the trajectory of these changes has been projected to deteriorate in the coming centuries (Adger *et al.*, 2003).

These extreme climate changes have a great toll on all aspects of human life, from food security, human health, to economic redundancy and poverty in most countries in the world. However, this impact is greatly felt in Africa, as these extreme weather conditions obstruct agricultural activities which is the keen contributor to the Gross Domestic Product of most African countries' economies. Furthermore, there exist a causal relationship between weather conditions and production variables. (Bryan *et al.*, 2009; Ochieng *et al.*, 2016). Numerous methodologies and models have been used in the literature to model variations in temperature. A few of these techniques are Machine Learning Techniques (random forests or neural networks) (Chidzalo *et al.*, 2022) and Time Series Analysis (ARIMA), Agent-Based Modeling, Probabilistic Graphical Models (GARCH) (Zmuk and Kovac, 2020).

Mean-reverting Itô diffusions driven by a standard Brownian motion with a time-dependent variance were used by Dornier and Querel (2000) to characterize temperature dynamics. Their study utilized only constant specifications into account when analyzing Chicago temperature data, and Monte Carlo simulation was then used. Due to the fact that, the OU model proposed by Dornier and Querel (2000) did not account for potential time-dependencies in the residuals observed from the regression model, Brody *et al.* (2002) considered a twist and proposed an OU process driven by a fractional Brownian motion based on a data series of daily temperatures from central England recorded from 1772 up to 1999, however, in the context of the Norwegian temperature data, Benth and Saltyte-Benth (2005) contended that the fractional Brownian dynamics model does not appear to be suitable.

Benth & Saltyte-Benth (2005) proposed an Ornstein–Uhlenbeck model with seasonal mean and volatility, in which the class of generalized hyperbolic Levy processes—a flexible class of Levy processes capturing the semi-heavy tails and skewness observed in some of the Norwegian temperature data—is used to generate the residuals rather than a Brownian motion. This rejection of the normal hypothesis was evident in various locations within the Norwegian temperature data. Gyamerah *et al.* (2018) used this model further to fit the daily average temperature from the northern region of Ghana. However, one approach that has been proven to best model temperature variations due to its mean-reverting properties in the context of Stochastic Differential Equations (SDEs) is Stochastic Differential Modeling, also known as Stochastic Temperature Models or Temperature stochastic Modelling. (Zahrnhofer, 2009; Zmuk & Kovac, 2020). By adding fluctuations and randomness to the modelling process, stochastic differential equations (SDEs) are essential to weather modelling. The intrinsically stochastic character of temperature fluctuations is captured in temperature modelling using SDEs, particularly in dynamic systems such as atmospheric processes or buildings. Stochastic Differential equations (SDEs) that describe temperature dynamics can be made more realistic by including stochastic variables. This allows for the modelling of random fluctuations and uncertainties that affect temperature changes over time. Stochastic temperature models have been proposed in literature by several studies that fit existing temperature models to daily average temperature observations to temperature data in several different countries, including Australia (Zahrnhofer, 2009), Canada (Swishchuk and Cui, 2012), and Morocco (Mraoua and Bari, 2007). Since various entities are impacted by weather in different ways. It is keenly important to fit these existing stochastic temperature model to Ghanaian data in order to understand the temperature dynamics.

The current paper adopts the Bhowan (2003) modified OU model in modeling the daily average temperature of the Bono region by assuming the volatility of the temperature process is a stochastic process which varies in on monthly basis but is nearly constant in a month. By filling in this knowledge void, stakeholders, farmers, insurers and financial institutions will be better equipped to build resilient agricultural risk systems that can withstand evolving environmental circumstances and improve our understanding of how weather changes and further develop mechanisms that captures its effect on national crop yield variability.

## II. MATERIALS AND METHOD

This section describes the various methodologies and techniques utilized in this paper to achieve the targeted results and findings.

### A. Weather Data Description.

This section of the paper examines the temperature data set, which was sourced from the Earth Observation and Innovation Centre (EORIC) of the University of Energy and Natural Resources, Sunyani Weather station located in Sunyani in the Bono Region of Ghana. The data collection, which yields 8,766 observations, included the daily average temperature, the quantity of rainfall, and the daily temperature difference for the period of January 1, 2000, to December 31, 2023 at two meters. Leap days entries were included in this data set, but they were not considered for the data analysis because of simplicity. Moreover, the dataset has no missing values.

We would perform simulation studies to project data regarding our model's quality. The Anderson-Darling test would be used to determine whether the data were normal.

### B. Construction of Temperature Model.

#### 1. Ornstein–Uhlenbeck model

The OU model was first utilized in temperature modeling by Dornier and Querel (2000) characterized temperature dynamics by mean-reverting Itô diffusions, driven by a standard Brownian motion given as;

$$dT(t) = d\tilde{T}(t) + \kappa_T \left(T(t) - \tilde{T}(t)\right)dt + \sigma_T(t)dB(t) \quad (1)$$

with

$$\tilde{T}(t) = a_T + b_T t + c_T \sin\left(\frac{2\pi t}{365} + \psi\right) \quad (2)$$

where $\tilde{T}(t)$ is known as the seasonal or mean function which describes the mean seasonal variation and sometimes referred to as the annual cycle or seasonality of the temperature, $\kappa_T > 0$ is the speed or rate of reversion which depicts the speed with which the temperature reverts to its mean, $\sigma_T(t) > 0$ is the volatility function.

Considering pattern of the temperature data in Figure 2, Equation (1) would be a good fit to the daily average temperature of the Bono Region of Ghana. However, we make an assumption that the temperature volatility alternates stochastically on a monthly basis, however, nearly constant during one month.

According to Zahrnhofer (2009) and Benth et al. (2007), the closed form solution to Equation (1) is given as;

$$T(t) = \tilde{T}(t) + \left(T(0) - \tilde{T}(0)\right)e^{-\kappa_T t} + \int_0^t \sigma_T(u)e^{-\kappa_T(t-u)}dB(u) \quad (3)$$

with an expectation given as;

$$E[T(t)] = \tilde{T}(t) + \left(T(0) - \tilde{T}(0)\right)e^{-\kappa_T t} \quad (4)$$

however, from Proposition 4.2 of Bhowan (2003), it is known that the expectation of $T(t)$ is given as;

$$E[T(t)] = \tilde{T}(t)$$

The parameters in Equation (1) would be estimated by discretizing Equation (3) using the Euler-Maruyama approximation method and estimated as follows.

#### 2. Estimation of Seasonal or Mean Function

Similarly, the parameters in the seasonal or mean function in Equation (2) are estimated as follows using the ordinary least squares estimation (OLSE). It should be noted that from Equation (2), $a_T$ is the mean level, $c_T$ is the amplitude level of the mean, $\psi$ is the phase angle and the period of oscillations is assumed to be one year neglecting leap year gives the weight $\frac{2\pi t}{365}$ (Swishchuk and Cui, 2012).

From Equation (3), $\tilde{T}(t)$ can be expressed as

$$\tilde{T}(t) = a_T + b_T t + c_T[\sin\left(\frac{2\pi t}{365}\right)\cos(\psi)]$$
$$+ c_T[\cos\left(\frac{2\pi t}{365}\right)\sin(\psi)]$$

$$\tilde{T}(t) = a_T + b_T t + [c_T \cos(\psi)]\sin\left(\frac{2\pi t}{365}\right)$$
$$+ [c_T \sin(\psi)]\cos\left(\frac{2\pi t}{365}\right)$$

$$y = \beta_0 + \beta_1 x_1 + \beta_2 x_2 + \beta_3 x_3 \quad (5)$$

where, $\beta_0 = a_T$, $\beta_1 = b_T$, $\beta_2 = c_T \cos(\psi)$, $\beta_3 = c_T \sin(\psi)$, $x_1 = t$, $x_2 = \sin\left(\frac{2\pi t}{365}\right)$, $x_3 = \cos\left(\frac{2\pi t}{365}\right)$

such as that $c = \frac{\beta_2}{\cos(\psi)}$ and $\psi = \tan^{-1}\left(\frac{\beta_2}{\beta_1}\right)$

#### 3. Estimation of Mean-Reverting Parameter

According to Bibby and Sørensen (1993), Bhowan (2003) and Zahrnhofer (2009) the mean reverting parameter is estimated using deriving the unbiased estimator of $\kappa_T$ which is the zero of the equation;

$$G_n(\kappa_T) = \sum_{j=1}^n \frac{\ddot{c}(T(j-1);\kappa_T)}{\sigma^2_T(j-1)}(T(j) - E[T(j)|T(j-1)]) \quad (6)$$

Where, $\ddot{c}(T(j-1);\kappa_T)$ denotes $\frac{\partial c}{\partial \kappa_T}$ and $n$ is equally the number of observations. The solution to Equation (5) is obtain by initially deriving $E[T(j)|T(j-1)]$ which is the conditional expectation of $T(j)$ given $T(j-1)$ which is obtained from the $E[T(t)]$ from Equation (3) as;

$$E[T(j)|T(j-1)] = \tilde{T}(i) + \left(T(j-1) - \tilde{T}(j-1)\right)e^{-\kappa_T} \quad (7)$$

for $s < t$

Substituting Equation (7) into (6) yields the estimate of $\kappa_T$ is given as

$$\hat{\kappa}_T = -\log\left[\frac{\sum_{j=1}^{n}\left(\frac{\tilde{T}(j-1)-T(j-1)}{\sigma^2_T(j-1)}[T(j)-\tilde{T}(j)]\right)}{\sum_{j=1}^{n}\left(\frac{\tilde{T}(j-1)-T(j-1)}{\sigma^2_T(j-1)}[T(j-1)-\tilde{T}(j-1)]\right)}\right] \quad (8)$$

It should be noted that $\sigma^2_T(j)$ refers to the calculated monthly volatilities as computed from Equation (9) below.

### 4. Estimation of Seasonal or Mean Function

For simplicity, one of the major assumption of this study is that, the volatility term $\sigma_T(t)$ is assumed to be a stochastic process, that is, the volatility changes on a monthly basis but it still constant during one month (Zahrnhofer, 2009).

Using quadratic variation, the SDE for the mean-reverting volatility is given as;

$$d\sigma_T(\rho) = \kappa_{\sigma_T}\left(\tilde{T}_{\sigma_T} - \sigma_T(\rho)\right)dr + \sigma_{T_{\sigma_T}}dB_{\sigma_T}(\rho) \quad (9)$$

where, $\tilde{T}_{\sigma_T} > 0$, is the trend or long-term volatility associated with the observed monthly temperature volatility process. From (10), $\sigma_{T_{\sigma_T}}$ is the volatility associated with the volatility process is estimated using the quadratic variation

$$\widehat{\sigma_{T_{\sigma_T}}^2} = \frac{1}{n}\sum_{h=0}^{n-1}(\sigma_T(h+1) - \sigma_T(h))^2 \quad (10)$$

Where $\sigma_T(h)$ is the quadratic variation of the tempertaure during a particular month $h$.

Lastly, $\kappa_{\sigma_T}$ is the rate of mean reversion for the volatility process is estimated similarly as Equation (6) and (7) using Equation (11)

$$\widehat{\kappa_{\sigma_T}} = -\log\left[\frac{\sum_{j=1}^{n}\left(\frac{\tilde{T}_{\sigma_T}-\sigma(j-1)}{\sigma^2_{\sigma_T}(j-1)}[\sigma(j)-\tilde{T}_{\sigma_T}]\right)}{\sum_{j=1}^{n}\left(\frac{\tilde{T}_{\sigma_T}-\sigma(j-1)}{\sigma^2_{\sigma_T}(j-1)}[\sigma(j-1)-\tilde{T}_{\sigma_T}]\right)}\right] \quad (11)$$

### 5. Simulation Studies

In visualizing the simulated sample paths of temperature based on our model, Equations (1) and (9) are discretized using the Euler-Maruyama approximation method which are given in Equations (12) and (13). For a particular month $\sigma_T(n)$ is simulated using Equation (13) and the simulated volatility process is then used in Equation (12) for the entire month's simulation

$$\Delta T(t_j) = \Delta\tilde{T}(t_j) + \kappa_T\left(T(t_j) - \tilde{T}(t_j)\right)\Delta t + \sigma_T(n)Z_j \quad (12)$$

where, $\Delta T(t_j) = T(t_{j+1}) - T(t_j)$ and $\Delta\tilde{T}(t_j) = \tilde{T}(t_{j+1}) - \tilde{T}(t_j)$

$$\sigma_T(n) - \sigma_T(n-1) = \kappa_{\sigma_T}\left(\tilde{T}_{\sigma_T} - \sigma_T(n-1)\right)dr + \sigma_{T_{\sigma_T}}Z_h \quad (13)$$

It should be noted that $Z_j$ and $Z_h \sim \mathcal{N}(0,1)$

## III. RESULT AND DISCUSSION

### A. Descriptive Statistics

The descriptive statistic parameters related to the measurement station (EORIC-Sunyani) is listed in Table 1. Conventionally, there are regional variations in the mean, median, minimum, and maximum daily mean temperature values in Ghana; nevertheless, these variations can be attributed to unique and widely dispersed geographical regions. Standard deviations (std) vary by location, the region experiences variation of 1.77°C and 5.36mm of temperature and precipitation respectively. The empirical distributions have negative kurtosis and an asymmetric form (skewness values that deviate from zero); in the instance of the Bono region, we see a unimodal pattern, which is mistakably indicative of significant seasonality. Again, the histogram of the daily average temperature for the Bono region as shown in Figure 1 below indicates that the data is not normally distributed. There appears to be a stronger concentration of probability around the mean of 26.08 than predicted by the normal distribution. Also, there is a right tail in the data, which is evident in Table 1 below, with a skewness of 0.51 with a kurtosis of -0.12). As confirmed in Figure 1b, this result depicts that the region experiences relatively greater dry seasons (harmattan) than wet seasons in the middle belt of Ghana; however, the Bono region experience warm temperatures and maximum precipitation up to 32.67°C and 126.51mm respectively.

Nevertheless, the Anderson-Darling test confirms the empirical distribution in Figure 1a and conclude that the normality test is rejected, as shown in Table 1. Hence, it may be argued that the temperature process cannot be driven by a Brownian motion but by other non-normal models such as the Lévy-driven process proposed by Benth and Saltyte-Benth (2005) or the fractional Brownian process proposed by Brody *et al.*, (2002). However, the daily temperature residuals as shown in Figure 4b, indicates some level of normality. Moreover, this study does not make much emphasis on the specific distribution of the temperature data, but rather seeks to focus on capturing the dynamics and variability of the temperature data using SDEs in our case the OU process. Due to that fact these models can incorporate stochastic process to account for the randomness and uncertainty, without necessarily relying on the assumptions about the underlying distribution of the temperature data.

In regard, we assume that the temperature process can be driven by a Brownian motion such as the OU process, particularly the modified OU process proposed by Bhowan (2003). For this study, OU models are modelled with Ghanaian data in order to ascertain an SDE model which captures the dynamics of temperature and rainfall for Ghanaian temperature and precipitation (rainfall) data based on some model evaluation metrics such as MSE, RSME, MAPE and the R squared.

Table 1. Descriptive Statistics for the daily average temperature and precipitation for Bono Region

|  | Temperature (°C) | Precipitation (mm) |  | Temperature (°C) | Precipitation (mm) |
|---|---|---|---|---|---|
| Mean | 26.08 | 3.60 | Max | 32.67 | 126.51 |
| Median | 25.75 | 2.120 | Min | 19.75 | 0.00 |
| SD | 1.77 | 5.36 | Skew | 0.51 | 6.21 |
| Kurtosis | -0.12 | 80.02 | $A^2$ statistic | 76.797 | 723.16 |
| p-value (5%)[two-tailed] | <0.001 | <0.001 |  |  |  |

*Source: Researcher's field data (2024)*

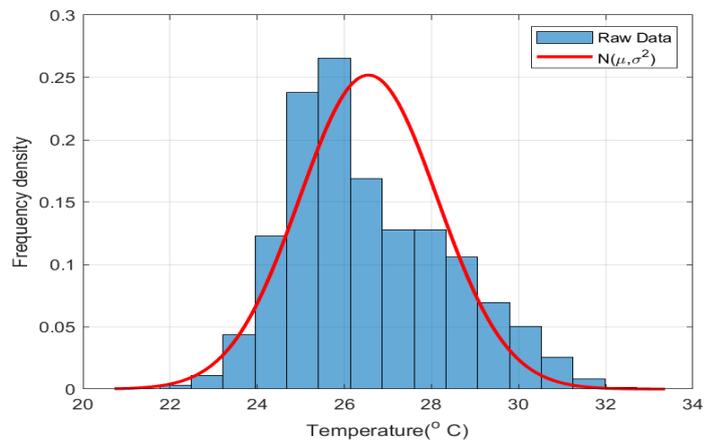

Figure 1a. Histogram of daily average temperature from the Bono Region of Ghana in the period from January 1, 2000 until December 31, 2023.

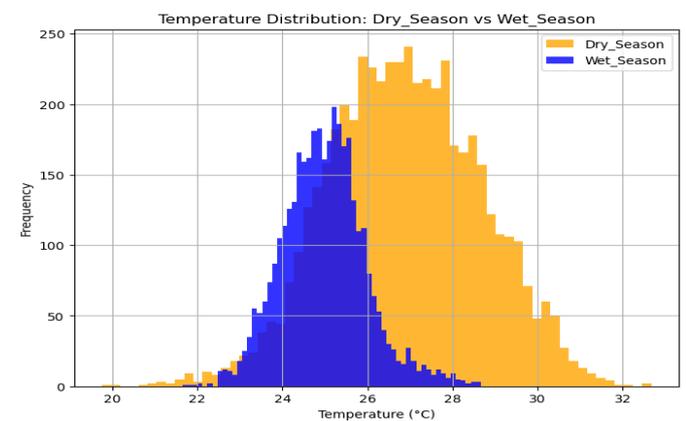

Figure 1b. Histogram of daily temperature distribution for Wet and Dry Seasons in the Bono Region of Ghana in the period from January 1, 2000 until December 31, 2023.

*B. Construction of Temperature Model.*

The observed daily average temperature of the Bono Region of Ghana spanning from January 1, 2000 to December 31, 2023 is depicted in Figure 2. It can be observed that, the Daily Average Temperature (DAT) of the region reverts about a temperature of 26 °C with maximum and minimum temperatures of 32.67°C and 19.75°C respectively.

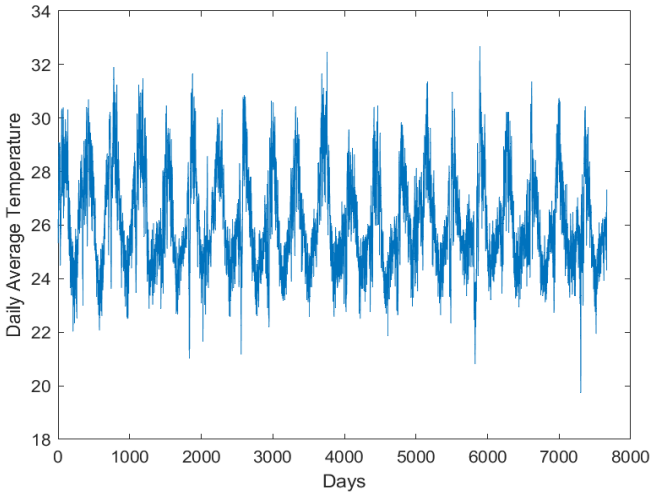

Figure 2. The observed daily average temperature from the Bono Region of Ghana from January 1, 2000, until December 31, 2020.
Source: Researcher's field data (2024)

### 1. Estimation of Seasonal or Mean Function

As depicted in Table 2, the estimated parameters of the seasonal mean or the annual mean is presented. Temperature is a variable that exhibits both regular periodic changes with a period of 365 units and steady change over time. The mean function $\tilde{T}(t)$ models temperature by combining a sinusoidal wave, a slow linear trend, and a constant value. The function can be interpreted as, a normal rainy season day and a dry season day differ in temperature by roughly 0.9°C, according to the sine function's amplitude of 1.75°C. Although the trend seems to be extremely slight, over a 24-year period it will indicate a mean temperature decrease of almost 0.7°C. Figure. 3 displays a visualization of this function together with the temperature data with an R squared coefficient of 51% indicating level of variability explained by the fitted seasonal mean or mean function. The average daily temperature for Bono Region and its estimated seasonal component $\tilde{T}(t)$ is visualized in Figure 3. This clearly illustrates that this basic parametrization can well capture the seasonal fluctuations in the daily average temperatures.

Table 2. Estimated Parameters of the Seasonal or Mean Function

| $\tilde{T}(t)$ | **Estimates** |
|---|---|
| $a_T$ | 26.4 |
| $b_T$ | -7.58e-05 |
| $c_T$ | 1.75 |
| $\psi_T$ | 0.531 |
| $R^2$ | 0.5062 |

*Source: Researcher's field data (2024)*

The fitted seasonal or Mean function is given as;

$$\tilde{T}(t) = 26.4 - 0.0000758t + 1.75\sin\left(\frac{2\pi t}{365} + 0.531\right) \quad (14)$$

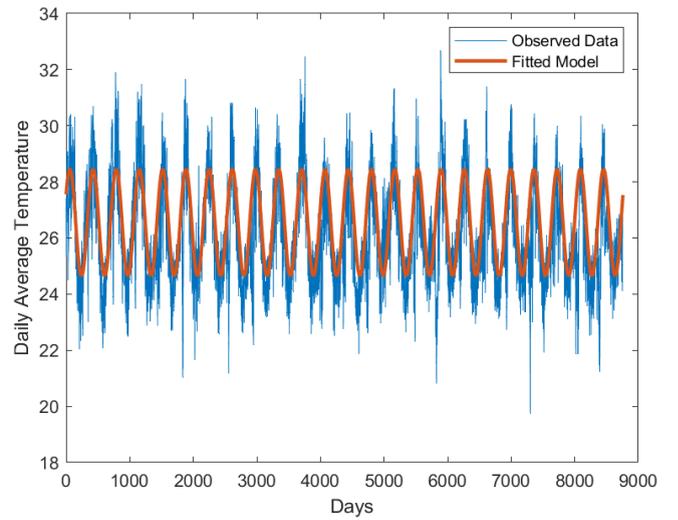

Figure 3. The observed daily average temperature from the Bono Region of Ghana from January 1, 2000, until December 31, 2020, together with the fitted seasonal or mean temperature function.

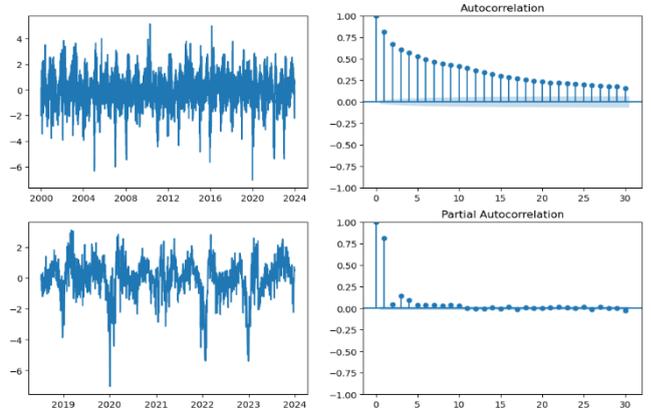

Figure 4a. The observed temperature residuals after de-trending and removing seasonality.

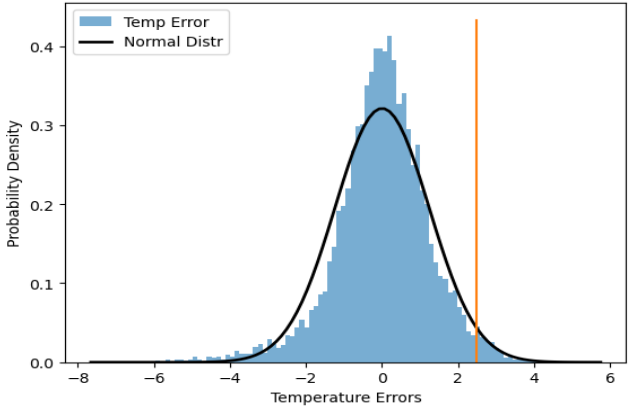

Figure 4b. Histogram of observed temperature residuals after de-trending and removing seasonality. for Bono Region in the period from January 1, 2000 until December 31, 2023.

Source: Researcher's field data (2024)

## 2. Estimation of Seasonal Volatility

The seasonal volatility constitutes what we refer to as the seasonal pattern in the residual. Seasonal volatility fundamentally makes sense. There is less variation in temperature during the dry season and more volatility during the wet season (Swishchuk and Cui, 2020).

However for simplicity, one of the major assumption of this study is that, the volatility term $\sigma_T(t)$ is assumed to be a stochastic process, that is, the volatility changes on a monthly basis but it still constant during one month (Zahrnhofer, 2009).

Using quadratic variation, the SDE for the mean-reverting volatility given in Equation (9). The $\tilde{T}_{\sigma_T} > 0$, is the trend or long-term volatility associated with the observed monthly temperature volatility process and its estimate as shown in Figure 5a and the estimate is given as;

$$\widehat{\tilde{T}_{\sigma_T}} = 0.877$$

From Equations (10) and (11), $\sigma_{T_{\sigma_T}}$, the volatility associated with the volatility process and $\kappa_{\sigma_T}$, the rate of mean reversion for the volatility process resopectively. The parameter estimates of the temperature volatility process is summarized in Table 3.

Table 3. Estimated Parameters of the mean-reverting stochastic volatility Function $\sigma_T(t)$

| $\sigma_T(t)$ | Estimates |
|---|---|
| $\widehat{\tilde{T}_{\sigma_T}}$ | 0.877 |
| $\widehat{\sigma_{\sigma_T}}$ | 0.419 |
| $\widehat{\kappa_{\sigma_T}}$ | 0.9890 |

*Source: Researcher's field data (2024)*

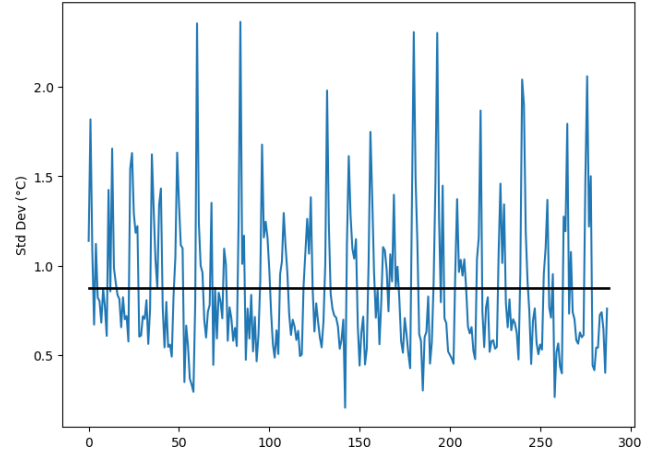

Figure 5a. The observed monthly volatility and $\tilde{T}_{\sigma_T}$ for Bono Region in the period from January 1, 2000 until December 31, 2023.

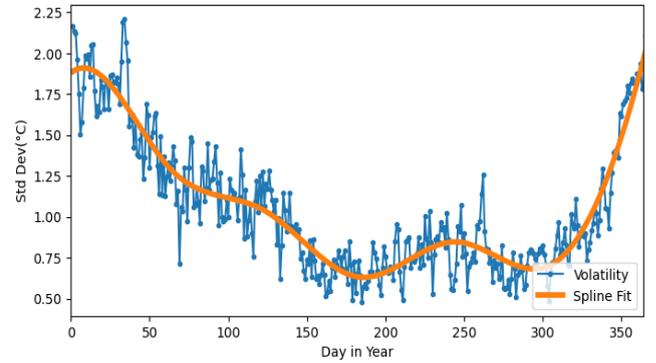

Figure 5b. The observed daily volatility with a Spline Fit for Bono Region in the period from January 1, 2000 until December 31, 2023.

Source: Researcher's field data (2024)

## 3. Estimation of Mean-Reverting Parameter

Solving Equation (8) numerically, the estimate of κ is given as

$$\hat{\kappa}_T = 0.1872$$

This shows that the speed or rate of mean reversion of the proposed temperature model is 0.1872. The mean reversion rate of 0.1872 in the temperature model tells us how quickly temperatures bounce back to the average level of about 26°C after reversion. This rate implies that around 18.72% of the difference from the average is adjusted each period (365 days). With this moderate rate, it shows that while temperature naturally tend to move back toward its usual range, it doesn't happen all at once; it takes some time for the effects of unusually warm or cold days to settle down and for temperature to get closer to it normal levels again as shown in Figure 6.

*4. Simulation Studies*

A simulation study was conducted to ascertain the efficacy of our model with simulated data. A 1000 simulated paths with a mean path of the daily average temperature of Bono Region from January 1, 2000, to December 31, 2023, was plotted based on the parameter estimates for our model in Equation (1). MATLAB R2022b was used in simulating and plotting the simulated paths as depicted in Figure 6. The corresponding evaluation metrics from this simulation are presented in Table 4 below.

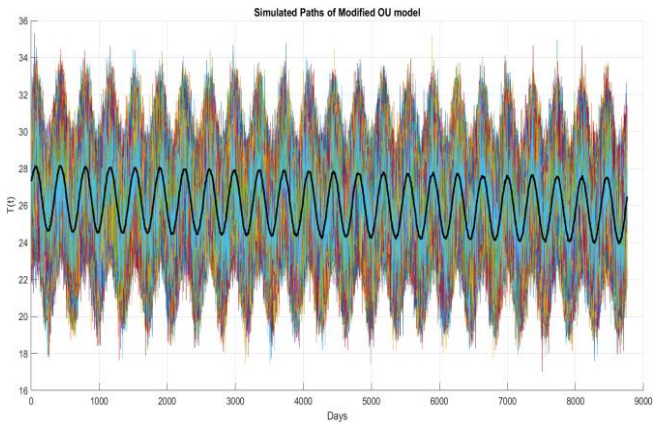

Figure 6. The 1000 simulation paths of the daily average temperature of the Bono Region of Ghana from January 1, 2000, until December 31, 2023, with the mean path.

As presented in Table 4 below, the evaluation metrics used in assessing our model's fitness to the daily average temperature of the Bono region were, the root mean squared error (RMSE), the mean absolute percentage error (MAPE) and the R squared coefficient are 1.2482, 3.5922% and 0.50182, which indicates that our model explains approximately 50% of the variations in the daily average temperature data of Bono Region. Statistically, the modified OU model proposed by Bhowan (2003) is a good model for the Bono Region's daily average temperature in Ghana.

Table 4. Model Evaluation Metrics for Temperature Model

|  | **Estimates** |
|---|---|
| RMSE | 1.2482 |
| MAPE | 3.5922% |
| $R^2$ | 0.50182 |

*Source: Researcher's field data (2024)*

## IV.  CONCLUSION

This paper sought to model the daily average temperature of the Bono region of Ghana stochastically, by adopting the modified Ornstein Uhlenbeck model proposed by Bhowan (2003) which as a modification of Alaton (2002) assumed that the volatility of the process was stochastic that is, the volatility varies randomly monthly but constant during one month. The novelty of the paper justifies the need for a weather (temperature) model based on Ghanaian data especially the Bono Region which is arguably one of the regions which feeds the whole country.

The results revealed that the Bono region experiences warm temperatures and maximum precipitation up to 32.67°C and 126.51mm respectively. It was observed that the mean Daily Average Temperature (DAT) of the region reverts about a temperature of 26 °C with maximum and minimum temperatures of 32.67°C and 19.75°C respectively. This can be interpreted as although the region is in the middle belt of Ghana, it still experiences warm(hot) temperatures daily and experiences dry seasons relatively more than wet seasons in the number of years considered for our analysis. Furthermore, it was observed that the trend or long-term volatility associated with the observed monthly temperature volatility process was 0.877 which reverts at a rate of 0.989. *Moreover, the seasonal or mean function revealed that* the difference between a normal rainy season day and a dry season day is approximately 0.9°C, according to the sine function's amplitude of 1.75°C. Although the trend seems to be extremely slight, over 24 years it will indicate a mean temperature decrease of almost 0.7°C. Furthermore, the rate of mean reversion was 0.1872 and this rate implies that around 18.72% of the difference from the average is adjusted each period (365 days). With this moderate rate, it shows that while temperature naturally tend to move back toward its usual rangeIt should be noted that the findings of this paper are relevant in the pricing of weather derivatives with temperature as an underlying variable in the Ghanaian financial and agricultural sector. Furthermore, it would assist in the development and design of tailored agriculture/crop insurance models which would incorporate temperature dynamics rather than extreme weather conditions/events such as floods, drought and wildfires.

This paper has provided fundamental information on the dynamics of the daily average temperature of the Bono Region of Ghana. This information reveals that the region being in the middle belt of Ghana still experiences warmer temperatures than cold temperatures and greater dry seasons than wet seasons hence it is recommended that farmers should plant crops and plantations which can thrive in the temperatures of 19.75°C - 32.67°C. Farmers should employ irrigation mechanisms in their farms which would help sustain the crops/plantations in the dry seasons. Since stochastic modelling of weather derivatives is relatively new in Ghana, further studies can be conducted into the modelling of more than one weather variable, other variables such as rainfall, humidity, and solar radiation amongst others thereby developing bivariate as well as tri-variate models which would capture the dynamics of weather in Ghana. Furthermore, future studies can investigate how other SDEs such as the CIR (Cox *et al.,* 1985) and Vasciek (Vasicek, 1977) models can be used to model weather and further pricing temperature derivatives.

## V. ACKNOWLEDGEMENT

The author would like to acknowledge the Earth Observation and Innovations Center (EORIC) of the University of Energy and Natural Resources, Sunyani for providing the weather data from their weather station. The author appreciates all reviewers for their insightful contributions to this paper.

## VI. CONFLICT OF INTEREST

It should be noted that, the author declares no conflict of interest.